\documentclass{article}

\usepackage{arxiv}

\usepackage[utf8]{inputenc} 
\usepackage[T1]{fontenc}    
\usepackage{hyperref}       
\usepackage{url}            
\usepackage{booktabs}       
\usepackage{amsfonts}       
\usepackage{nicefrac}       
\usepackage{microtype}      
\usepackage{lipsum}

\title{Black-Hole evaporation and Quantum-depletion in Bose-Einstein condensates}

\author{
 Ivan Arraut\\
  The Open University of Hong Kong\\
  30 Good Shepherd Street, Homantin, Kowloon\\
  \texttt{ivanarraut05@gmail.com} 
}

\begin{document}
\maketitle

\begin{abstract}
We study the analogy between the Hawking radiation in Black-Holes and the Quantum depletion process of a Bose-Einstein condensate by using the Bogoliubov transformations method. We find that the relation between the Bogoliubov coefficients is similar in both cases (in the appropriate regimes). We then connect the condensate variables with those associated to the Black-Hole, demonstrating then that the zero temperature regime of the condensate is equivalent to the existence of an event horizon in gravity.  
\end{abstract}


\section{Introduction}

The phenomena of Quantum depletion in Bose condensation was observed recently experimentally \cite{1}. The analogy of this phenomena and Black-Holes suggests that the Quantum depletion is just equivalent to the Hawking radiation \cite{2}. Indeed, it was discovered by Hawking before that the Black-Holes not only absorb particles but they are also able to emit them. By then, Hawking used arguments of Quantum Mechanics over a classical background (semi-classical) \cite{3}. Hawking argument compares the modes of a scalar field defined in the past null infinity with those modes defined in the future null infinity for the same field. It comes out that the modes in the future null infinity do not contain the whole information of the scalar field. There are some additional modes completing the information of the scalar field, but they only have data at the event horizon of the Black-Hole and their information is not available for the observers. The main point is that the unitarity is lost when we compare the modes defined in the future null infinity with the modes defined in the past null infinity. Such comparison has to be done in a single vacuum which is defined by convention at the past null infinity which is devoid of particles \cite{3}. Several variations to the Hawking result have been explored in different scenarios. In \cite{Var1}, the Hawking radiation in a General cubic (curvature) theory of gravity was explored. In \cite{Var2}, the effects of the Hawking radiation in a two-level atom were analyzed. In \cite{Var3}, oscillating bounces in the Schwarzschild de-Sitter space were investigated. Such effects might affect the rate of evaporation of the Black-Hole. In \cite{Var4}, the grey-body factor effects and how they affect the rate of evaporation were studied. In \cite{Var5}, different effects inside the scenario of gravity's rainbow were explored. In such scenarios, Black-Hole remnants as well as modifications of the dispersion relations and the Black-Hole thermodynamics appear naturally as a consequence of the rainbow effect. In \cite{Var6, Var61}, the quantum gravity effects appearing in the form of a Generalized Uncertainty Principle (GUP), were considered in order to analyze deformations in the Black-Hole thermodynamics and the Hawking radiation. The effect of thermal fluctuations in the Black-Hole thermodynamics was considered in \cite{Var7, Var71}. The Black-Hole thermodynamics in modified gravity theories was analyzed in \cite{Var71, Var8}. In \cite{Var9}, the Black-Hole information paradox was explored, this paradox has a direct connection with the Hawking radiation. In \cite{Var10}, the stability of some Black-Hole solutions in some modified gravity theories were explored. The stability of Black-Holes is fundamental for any subsequent analysis, including of course the Black-Hole evaporation. Virtual Black-Holes were proposed in \cite{Var11} in order to explore possible modifications appearing in the definition of Black-Hole entropy. Interesting scenarios, connecting thermodynamic properties of the Black-Hole with condensed matter systems via holographic principle and related Quantum gravity proposals were analyzed in \cite{Var12}. The Hawking radiation for the asymptotically de-Sitter and Anti de-Sitter solutions was studied in \cite{Var61, Var13}. Non-commutative scenarios, exploring modifications of the Black-Hole thermodynamics, were studied in \cite{Var14}. In \cite{Var15}, Quantum gravity candidates, with different Lifshitz scaling for time and space coordinate were also explored. The connection between the Black-Hole evaporation and the phenomena of Quantum-depletion of Black-Holes was proposed by Dvali et al in \cite{2} by using the Black-Hole N-portrait which makes a connection between the size of the system (event horizon) and the number of gravitons involved in the condensate. From this perspective the Black-Hole is a leaky bound state of a condensate of $N$ interacting soft gravitons existing in a critical condition. The analogy was extended to the scenario of neural networks where the degrees of freedom of a gas of bosons were mapped to the degrees of freedom representing a neural network arrangement \cite{4}. The analysis of the Black-Hole evaporation from the perspective of neural networks was done by the author in \cite{5} without considering the phenomena of Quantum depletion. In \cite{5}, the author just mapped the relation between the Bogoliubov coefficients, necessary for the reproduction of the Hawking radiation onto the modes expanded in the Hamiltonian representing a neural network arrangement. In this paper we analyze the conditions necessary for the process of Quantum depletion to be equivalent to the process of Black-Hole evaporation. The conditions are obtained by deriving the Bogoliubov coefficients independently for both situations and subsequently comparing the corresponding results. We then fins an expression for the surface gravity of the Black-Hole as a function of the variables of the condensate which include the momentum of the particles able to escape the condensate and the frequency of the associated modes. We focus on the low momentum regime since the particles escaping the condensate are supposed to have small momentum due to the fact that the condensate under analysis has a temperature near the absolute zeto $T=0$. Our results show the consistent relation suggesting that for a fixed momentum for the particles escaping the system, a lower frequency for the modes implies a larger mass associated to the Black-Hole.          

\section{Quantization of a gas of bosons and Quantum depletion}   \label{QDC}

Consider a gas of bosons represented by a Quantum field $\phi({\bf r})$ with the dynamic obeying the following Hamiltonian

\begin{equation}   \label{Ham}
\hat{H}=\frac{\hbar^2}{2m}\int d^3r\phi^+({\bf r})\nabla^2\phi({\bf r})+\frac{1}{2}\int\int\phi^+({\bf r}_1)\phi^+({\bf r}_2)u({\bf r}_1, {\bf r}_2)\phi({\bf r}_2)\phi({\bf r}_1).
\end{equation}
Here $u({\bf r}_1, {\bf r}_2)$ is the two-body interaction potential \cite{Pathria}. Since the Quantum depletion phenomena is expected to happen at low temperatures, namely, near the absolute zero; then the momentum of the particles at that point is expected to be small. We define the particle number operator as

\begin{equation}
\hat{N}=\int d^3r\psi^+({\bf r})\psi({\bf r}).    
\end{equation}
The particle number is a conserved quantity for the Hamiltonian under analysis and then its operator commutes with the Hamiltonian operator as

\begin{equation}
[\hat{N}, \hat{H}]=0.    
\end{equation}
If we make the expansion in a Fourier series of the Quantum field $\psi({\bf r})=\sum_i\hat{a}_iu_i({\bf r})$, together with its complex conjugate, then we get the standard definition of the particle number operator as

\begin{equation}
\hat{N}=\sum_i\hat{a}^+_i\hat{a}_i.    
\end{equation}
Here the summation is carried out over all the particle states of the system. Evidently, for the single particle state we can define the particle number operator as $\hat{N}_i=\hat{a}^+_i\hat{a}_i$ (omitting the summation over index). The Hamiltonian (\ref{Ham}) itself can be also expressed in terms of annihilation and creation operators, written in this form, it becomes

\begin{equation}   \label{sumpot}
\hat{H}=\sum_{\bf p}\frac{p^2}{2m}\hat{a}^+_{\bf p}\hat{a}_{\bf p}+\frac{1}{2}\sum_{{\bf p}_1', {\bf p}_2'}u^{{\bf p}_1', {\bf p}_2'}_{{\bf p}_1, {\bf p}_2}\hat{a}^+_{{\bf p}_1'}\hat{a}^+_{{\bf p}_2'}\hat{a}_{{\bf p}_2}\hat{a}_{{\bf p}_1}.    
\end{equation}
Here $u^{{\bf p}_1', {\bf p}_2'}_{{\bf p}_1, {\bf p}_2}$ is just a matrix element representing the interaction of two particles as it is defined in \cite{Pathria}. The transfer momentum due to the interaction of a pair of particles, is defined as

\begin{equation}
{\bf p}=({\bf p}_2-{\bf p}_2')=-({\bf p}_1-{\bf p}_1').    
\end{equation}
Evidently, the momentum conservation in the form ${\bf p}_1'+{\bf p}_2'={\bf p}_1+{\bf p}_2$ is satisfied. If the spin of the particles is considered, it can be also included in the Hamiltonian. Here for our purposes, it is irrelevant to include the spin of the particles. One relevant quantity to define is the scattering amplitude, here defined as

\begin{equation}
a({\bf p})=\frac{m}{4\pi\hbar^2}\int u({\bf r})e^{i{\bf p}\cdot{\bf r}/\hbar}d^3r.    
\end{equation}
At the lowest order in momentum, which is the case of low-energy scattering, we have the limiting result

\begin{equation}   \label{limit value}
a\approx\frac{mu_0}{4\pi\hbar^2},\;\;\;u_0=\int u({\bf r})d^3r.    
\end{equation}
\subsection{Low temperature behavior of a Bose gas}

At low temperatures, the matrix element $u^{{\bf p}_1', {\bf p}_2'}_{{\bf p}_1, {\bf p}_2}$, takes the limit value $u_0/V$ as it is defined in eq. (\ref{limit value}). Note that since in this case we have zero-momentum transfer ${\bf p}=0$, then in the sum for the potential term in eq. (\ref{sumpot}), we only consider those terms with a vanishing momentum transfer. In this way, the Hamiltonian becomes

\begin{equation}   \label{Hamiltonexpr}
\hat{H}=\sum_{\bf p}\frac{p^2}{2m}\hat{a}^+_{\bf p}\hat{a}_{\bf p}+\frac{2\pi a\hbar^2}{mV}\left(\sum_{\bf p}\hat{a}^+_{\bf p}\hat{a}^+_{\bf p}\hat{a}_{\bf p}\hat{a}_{\bf p}+\sum_{{\bf p}_1\neq{\bf p}_2}(\hat{a}^+_{\bf p_1}\hat{a}^+_{\bf p_2}\hat{a}_{\bf p_2}\hat{a}_{\bf p_1}+\hat{a}^+_{\bf p_2}\hat{a}^+_{\bf p_1}\hat{a}_{\bf p_2}\hat{a}_{\bf p_1})\right).   
\end{equation}
It can be easily proved that the energy eigenvalues for this Hamiltonian are defined as

\begin{equation}
E_{\bf p}\approx \sum_{\bf p}n_{\bf p}\frac{p^2}{2m}+\frac{2\pi a\hbar^2}{mV}(2N^2-n_0^2).      
\end{equation}
From this, it is clear that the ground state is defined as $E_0\approx 2\pi a\hbar^2N^2/mV$. 

\subsection{Quantum depletion}

At low temperatures, the emission of particles in the system is dominated by the phenomena of Quantum depletion. When the temperature is near the zero value, the particles are not supposed to escape. However, Quantum depletion still allows the possibility for this to happens. This is analogous to the fact that the particles are not supposed to escape from the event horizon of a Black-Hole. Indeed, later we will demonstrate that this fact is the key point for the comparison with the Hawking radiation. Near the zero temperature, it is expected for most of the particles in the system to be in the ground state where the particle number is $n_0$ with ${\bf p}=0$. In this way $n_0/N\approx1$. In the same form, the number of excited particles with ${\bf p}\neq0$, is considered to be small and then $n_{\bf p}/N<<1$. In this way, we can use the approximation    

\begin{equation}
2N^2-n_0^2\approx N^2+2N\sum_{{\bf p}\neq0}\hat{a}^+_{\bf p}\hat{a}_{\bf p},    
\end{equation}
which can be used for simplifying the Hamiltonian expression given in eq. (\ref{Hamiltonexpr}). In addition, due to the large amount of particles in the ground state, we can take the annihilation and creation operators of the found state as c-numbers, namely $\hat{a}_0\hat{a}_0^+=\hat{a}_0^+\hat{a}_0$. Considering this aspect and still using the zero-order approximation for $u({\bf r})$ given in eq. (\ref{limit value}), then we get a simplified version of the Hamiltonian given by

\begin{equation}   \label{impro}
\hat{H}=\sum_{\bf p}\frac{p^2}{2m}\hat{a}^+_{\bf p}\hat{a}_{\bf p}+\frac{u_0}{2V}\left(N^2+N\sum_{{\bf p}\neq0}(2\hat{a}^+_{\bf p}\hat{a}_{\bf p}+\hat{a}^+_{\bf p}\hat{a}_{-\bf p}^++\hat{a}_{\bf p}\hat{a}_{-\bf p})\right). 
\end{equation}
It can be proved that if we consider $u_0$ expanded up to second-order, considering then the transitions of the system, we obtain

\begin{equation}
u_0\approx \frac{4\pi a\hbar^2}{m}\left(1+\frac{4\pi a\hbar^2}{V}\sum_{{\bf p}\neq0}\frac{1}{p^2}\right).    
\end{equation}
In this way, the Hamiltonian (\ref{impro}), becomes

\begin{equation}   \label{Hamilton}
\hat{H}=\sum_{{\bf p}\neq0}\frac{p^2}{2m}\hat{a}^+_{\bf p}\hat{a}_{\bf p}+\frac{2\pi a\hbar^2}{m}\frac{N}{V}\sum_{{\bf p}\neq0}(2\hat{a}^+_{\bf p}\hat{a}_{\bf p}+\hat{a}^+_{\bf p}\hat{a}_{-\bf p}^++\hat{a}_{\bf p}\hat{a}_{-\bf p})+\frac{2\pi a\hbar^2}{m}\frac{N^2}{V}\left(1+\frac{4\pi a\hbar^2}{V}\sum_{{\bf p}\neq0}\frac{1}{p^2}\right).    
\end{equation}
The energy level of the system can be found after diagonalization of the Hamiltonian by using the Bogoliubov transformations in the form \cite{equi}

\begin{equation}   \label{TransformationsBogo}
\hat{b}_{\bf p}=\frac{\hat{a}_{\bf p}+\alpha_{\bf p}\hat{a}^+_{-{\bf p}}}{\sqrt{1-\alpha^2_{\bf p}}},\;\;\;\;\;\hat{b}^+_{\bf p}=\frac{\hat{a}^+_{\bf p}+\alpha_{\bf p}\hat{a}_{-{\bf p}}}{\sqrt{1-\alpha^2_{\bf p}}}.   
\end{equation}
These transformations can be inverted as 

\begin{equation}   \label{TransformationsBogo2}
\hat{a}_{\bf p}=\frac{\hat{b}_{\bf p}-\alpha_{\bf p}\hat{b}^+_{-{\bf p}}}{\sqrt{1-\alpha^2_{\bf p}}},\;\;\;\;\;\hat{a}^+_{\bf p}=\frac{\hat{b}^+_{\bf p}-\alpha_{\bf p}\hat{b}_{-{\bf p}}}{\sqrt{1-\alpha^2_{\bf p}}},
\end{equation}

with

\begin{equation}   \label{alpha}
\alpha_{\bf p}=\frac{mV}{4\pi a\hbar^2N}\left(\frac{4\pi a\hbar^2N}{mV}+\frac{p^2}{2m}-\epsilon({\bf p})\right).    
\end{equation}
Here $\epsilon({\bf p})$ represents the dispersion relation, which here is defined as

\begin{equation}   \label{Dispersion}
\epsilon({\bf p})=\left(\frac{4\pi a\hbar^2N}{mV}\frac{p^2}{m}+\left(\frac{p^2}{2m}\right)^2\right)^{1/2}.    
\end{equation}
Evidently, the commutation relations are unchanged after executing a Bogoliubov transformation. Then we have

\begin{equation}   \label{Commutation}
[\hat{a}_{\bf p}, \hat{a}_{\bf p'}]=[\hat{b}_{\bf p}, \hat{b}_{\bf p'}]=\delta_{{\bf p}, {\bf p'}},    
\end{equation}
with all the other commutators vanishing. At the ground state, since the operators $\hat{a}_{\bf p}$ behave as $c$-numbers, the previous commutator will naturally vanish in such situations. After using the transformations (\ref{TransformationsBogo}) inside the Hamiltonian (\ref{Hamilton}), then we get

\begin{equation}   \label{NewHamilton}
\hat{H}=E_0+\sum_{{\bf p}\neq0}\epsilon({\bf p})\hat{b}^+_{\bf p}\hat{b}_{\bf p}, 
\end{equation}
with $E_0$ defined as 

\begin{equation}
E_0=\frac{2\pi a\hbar^2N^2}{mV}+\frac{1}{2}\sum_{{\bf p}\neq0}\left(\epsilon({\bf p})-\frac{p^2}{2m}-\frac{4\pi a\hbar^2N}{mV}+\left(\frac{4\pi a\hbar^2N}{mV}\right)^2\frac{m}{p^2}\right).    
\end{equation}
The form of the Hamiltonian (\ref{NewHamilton}) together with the commutation relations (\ref{Commutation}), suggests that $\hat{b}^+_{\bf p}$ is the creation operator of some quasiparticle and $\hat{b}_{\bf p}$ corresponds to the annihilation operator of the same quasiparticle. Evidently $\hat{b}^+_{\bf p}\hat{b}_{\bf p}$ is the particle number operator of the quasi-particle. The dispersion relation for the quasi-particle is given by eq. (\ref{Dispersion}). The form of the Hamiltonian (\ref{NewHamilton}) suggests that $E_0$ is the ground state of the system and that the quasi-particles behave as Bosonic field. If the ground state of the system were perfect, we would have zero particles with non-vanishing momentum, in agreement with the eigenvalues $\hat{n}_{{\bf p}\neq0}^b=\hat{b}^+_{\bf p}\hat{b}_{\bf p}$ which would vanish in such a case. However, these eigenvalues do not represent the behavior of real particles. Additionally, in order to get a finite energy for the ground state, it is necessary to have some particles having non-zero energy even at absolute zero \cite{Pathria}. 

\subsection{Evaporation of Physical particles}   \label{Physicalpart}

The particle number corresponding to physical particles is defined by $\hat{n}_{\bf p}^a=\hat{a}^+_{\bf p}\hat{a}_{\bf p}$. We can then find the number of real particles perceived from the perspective of the vacuum defined by the quasi-particles as $<\bar{0}\vert\hat{n}_{\bf p}^b\vert\bar{0}>=0$. Then we find

\begin{equation}   \label{mama}
<\bar{0}\vert\hat{n}_{\bf p}^a\vert\bar{0}>=<\bar{0}\vert\hat{a}^+_{\bf p}\hat{a}_{\bf p}\vert\bar{0}>=\frac{\alpha_{\bf p}^2}{1-\alpha_{\bf p}^2}. 
\end{equation}
This result is obtained after replacing the Bogoliubov transformations (\ref{TransformationsBogo2}) inside the expression $<\bar{0}\vert\hat{a}^+_{\bf p}\hat{a}_{\bf p}\vert\bar{0}>$ in eq. (\ref{mama}). We can analyze further the role of the Bogoliubov coefficients in eq. (\ref{TransformationsBogo2}), if we express the Bogolibov transformation in general as

\begin{equation}   \label{realbogo}
\hat{a}_{\bf p}=u_{{\bf p}, {\bf p'}}\hat{b}_{\bf p'}-v_{{\bf p}, {\bf p'}}\hat{b}_{\bf p'}^+. 
\end{equation}
Then we can define

\begin{equation}   \label{losAlamos}
u_{{\bf p}, {\bf p'}}=\frac{1}{\sqrt{1-\alpha^2_{\bf p}}},\;\;\;v_{{\bf p}, {\bf p'}}=\frac{\alpha_{\bf p}}{\sqrt{1-\alpha^2_{\bf p}}}.    
\end{equation}
It is trivial to demonstrate that 

\begin{equation}
\vert u_{{\bf p}, {\bf p'}}\vert^2-\vert v_{{\bf p}, {\bf p'}}\vert^2=1.    
\end{equation}
This demonstrates that the commutation relations are preserved after doing the Bogoliubov transformations. From eq. (\ref{losAlamos}), it is important to notice the relation between the Bogoliubov coefficients as

\begin{equation}
v_{{\bf p}, {\bf p'}}=\alpha_{\bf p}u_{{\bf p}, {\bf p'}}.  
\end{equation}
This relation suggests that $\vert v_{{\bf p}, {\bf p'}}\vert=\alpha_{\bf p}\vert u_{{\bf p}, {\bf p'}}\vert$

\subsection{A comparison with the Black-Hole evaporation process}

The quantum depletion process is analogous to the Black-Hole evaporation process \cite{3}. Then in the same way as the particles are not supposed to escape from a condensate at zero temperature; no particle is supposed to escape from a Black-Hole classically \cite{BH}. The particles can escape from a Black-Hole via Hawking radiation in the same way as the particles can escape from a condensate at zero temperature via Quantum-depletion. In order to understand this fact, we need to analyze the Black-Hole evaporation process. The Hawking radiation process is a natural consequence of a comparison between the vacuum located at the future null infinity, namely, the vacuum with no-gravity and defined after the formation of the Black-Hole; with respect the vacuum located in the past null-infinity, namely, the one defined before the Black-Hole formation and devoid of particles. Naturally, before the Black-Hole formation (past-null infinity) there are no particles; then we can settle the vacuum in such a case as

\begin{equation}
\hat{b}_{\bf p}\vert\bar{0}>=0.    
\end{equation}
Then the vacuum defined in the past-null infinity is  consistent with the vacuum defined by the quasi-particles in the subsection (\ref{Physicalpart}). On the other hand, the vacuum defined at the future null infinity, is consistent with the vacuum defined for the real particles in the same subsection. Then we should expect the observers located at the future null-infinity to detect particles, even in these are not supposed to escape from the Black-Hole. The standard calculation done by Hawking in \cite{3}, suggests that a scalar field in the past-null infinity can be expanded as 

\begin{equation}   \label{field1}
\phi(x, t)=\sum_{\bf p}\left(f_{\bf p}\hat{b}_{\bf p}+\bar{f}_{\bf p}\hat{b}_{\bf p}^+\right).    
\end{equation}
The same scalar field, expanded at the future-null infinity, requires two different set of modes, then we can write

\begin{equation}   \label{field2}
\phi(x, t)=\sum_{\bf p}\left(p_{\bf p}\hat{a}_{\bf p}+\bar{p}_{\bf p}\hat{a}_{\bf p}^++q_{\bf p}\hat{c}_{\bf p}+\bar{q}_{\bf p}\hat{c}_{\bf p}^+\right).    
\end{equation}
Both equations, (\ref{field1}) and (\ref{field2}) carry the same information. The vacuum $\hat{a}_{\bf p}\vert0>=0$ is defined at the future null-infinity. On the other hand, the modes $q_{\bf p}$, together with the operators $\hat{c}$, are defined at the event horizon. These are the modes which the observer located at the asymptotic future cannot see, generating the famous information problem in Black-Holes \cite{3}. Here we focus only on the modes which are perceived by the observers. In order to relate the modes defined by $f_{\bf p}$ and those defined by $p_{\bf p}$, we employ the Bogoliubov transformations. Following the arguments of Hawking in \cite{3}, we can find that the relation between the modes under discussion, follow the same Bogoliubov transformations defined in eq. (\ref{realbogo}), and evidently

\begin{equation}   \label{mama2}
<\bar{0}\vert\hat{n}_{\bf p}^a\vert\bar{0}>=\vert v_{{\bf p}, {\bf p'}}\vert^2,
\end{equation}
which is related to the result (\ref{mama}). The arguments of Hawking demonstrated that the fraction of particles escaping the Black-Hole, follow the following distribution

\begin{equation}
<\bar{0}\vert\hat{n}_{\bf p}^a\vert\bar{0}>=\frac{\Gamma_{{\bf p}, {\bf p'}}}{e^{\frac{2\pi\omega}{\kappa}}-1},    
\end{equation}
if the particles escaping the Black-Hole are bosons. In the previous expression, $\Gamma_{{\bf p}, {\bf p'}}$ represents the fraction of particles entering the collapsing body (Black-Hole). One key ingredient in the Hawking calculation is the relation between the Bogoliubov coefficients in the following way

\begin{equation}
\vert u_{{\bf p}, {\bf p'}}\vert=e^{\frac{\pi\omega}{\kappa}}\vert v_{{\bf p}, {\bf p'}} \vert. 
\end{equation}
If we take the Quantum-depletion process as analogous to the Hawking radiation, then we conclude that $\alpha_{\bf p}=e^{\frac{\pi\omega}{\kappa}}$. Here, naturally $\omega$ is a function of ${\bf p}$. We can easily solve for the surface gravity, obtaining then the temperature associated to the depletion to be proportional to

\begin{equation}   \label{Impressiveyou are}
\kappa=\frac{\pi\omega}{Ln\vert\alpha_{\bf p}\vert}=\frac{\pi\omega}{Ln\frac{\vert u_{{\bf p}, {\bf p'}}\vert}{\vert v_{{\bf p}, {\bf p'}}\vert}}.    
\end{equation}
Since the Quantum-depletion occurs at low temperature, and the momentum associated to the particles is low, then we will focus on the low momentum limit. Before that, we can simplify $\alpha_{\bf p}$ with the help of eqns. (\ref{alpha}) and (\ref{Dispersion}) and using the change of variables 

\begin{equation}
x=p\left(\frac{V}{8\pi a\hbar^2N}\right)^{1/2}.    
\end{equation}
In this way, we obtain

\begin{equation}
\alpha_{\bf p}=1+x^2-\sqrt{2x^2+x^4}.
\end{equation}
The low momentum regime is equivalent to $x<<1$ and then we get the approximation

\begin{equation}
\alpha_{\bf p}\approx 1-\frac{3}{\sqrt{2}}x.    
\end{equation}
If we replace this expression in eq. (\ref{Impressiveyou are}), then we get the result 

\begin{equation}
\kappa\approx\frac{\pi\omega}{\sum_{m=1}^\infty\frac{(-1)^{m-1}}{m}\left(\frac{\sqrt{3}}{2}x\right)^m}.  
\end{equation}
If we expand up to the first order term in the logarithm expansion, then we get

\begin{equation}
\kappa\approx\frac{2\pi\omega}{\sqrt{3}x}.  
\end{equation}
We can then proceed to make some interpretations. Note that the previous expression connects the concept of surface gravity with the condensed matter variables associated to the modes. Hawking demonstrated that for Black-Holes, $\kappa=1/2\pi M$, then we can conclude that 

\begin{equation}
\frac{\omega}{x}=\frac{\sqrt{3}}{4\pi^2M}.    
\end{equation}
This result suggests that a large mass for the Black-Hole is associated to a low frequency for the modes of the condensate. Here it is always assumed that $x<<1$. The negative heat capacity associated to the Black-Holes, emerges then naturally from the Quantum depletion of the condensate as it was suggested in a similar scenario in \cite{2}. 

\section{Conclusions}

In this paper we have analyzed the similarities between the process of Hawking radiation in Black-Holes and the process of Quantum-depletion in Bose-Einstein condensates. We have demonstrated that the Bogoliubov coefficients obey similar relations in both situations. In this way, we could make a direct connection between the variables and parameters of the condensate with those related to the Black-Hole. From the perspective of Black-Holes, the existence of an event horizon implies that classically no particles can escape from a Black-Hole. In the same way, at zero temperature of the condensate, no particle is supposed to escape from it if we employ the classical approximation. In both situations, namely, Black-Holes and Bose-Einstein condensates, the Quantum effects (Hawking radiation or Quantum-depletion) allow the particles to escape. In the case of a condensate this gives space to the phenomena of Quantum depletion and for the case of Black-Holes, the Quantum effects give space to the phenomena of Hawking radiation. Further studies are necessary for analyzing possible connections with the holographic principle \cite{Malda}. One immediate problem to be explored is the gravitational side of the strongly correlated limit for some condensed matter systems of the form analysed in \cite{Varconc1}. The results connected to the phenomena of Quantum depletion for strongly correlated systems were obtained in \cite{Varconc2}. Mapping such results to the gravitational side by using the holographic principle is an open problem to be explored in a subsequent paper.          

{\bf Acknowledgement}
The author would like to thank Naoki Yamamoto for the hospitality during the visit to Keio University in Yokohama-Japan, where part of this work was presented.  

\bibliographystyle{unsrt}  


\end{document}